# Construction of Turbo Code Interleavers from 3-regular Hamiltonian Graphs

Arya Mazumdar, *Student Member, IEEE,* A K Chaturvedi, *Senior Member, IEEE,* Adrish Banerjee, *Member, IEEE*

*Abstract*—In this letter we present a new construction of interleavers for turbo codes from 3-regular Hamiltonian graphs. The interleavers can be generated using a few parameters, which can be selected in such a way that the girth of the interleaver graph (IG) becomes large, inducing a high summary distance. The size of the search space for these parameters is derived. The proposed interleavers themselves work as their de-interleavers.

## I. INTRODUCTION

The distance spectrum of a turbo code is influenced greatly by the interleaver used. In [1] a graphical representation of an interleaver was introduced and it was shown that good interleavers can be constructed from graphs having large girth. In this paper, we propose a new graph-based interleaver design technique from 3-regular Hamiltonian graphs that tries to maximize the summary distance of the interleaver by increasing the girth of the interleaver graph. To obtain 3-regular Hamiltonian graphs with large girths we use the construction of [2]. We determine the size of the search-space of the parameters needed for the construction. The proposed graph-based interleavers themselves work as their de-interleavers.

## II. DEFINITIONS AND PRELIMINARIES

Every vertex of a $k$-regular graph has degree $k$. The number of distinct edges in a cycle of a graph is called the length of the cycle. The length of the shortest cycle of a graph is the girth of that graph. A Hamiltonian cycle is a path that traverses every vertex exactly once and comes back to the origin.

The Lee distance metric on the set $\{0, 1, 2, ....., N-1\}$ is defined in [3] as,

$$d_L(i,j) = \min(|i-j|, N-|i-j|), \ 0 \le i, j \le N-1 \quad (1)$$

A 'cycle in the permutation $\pi$' [3], for any even $l$, is defined as the sequence $\sigma = i_1, i_2, j_2, j_3, i_3, i_4, ..., j_l, j_1$ where $0 \le i_k, j_k \le N-1$, such that all $i_k$ are different and $j_k = \pi(i_k)$, for $k = 1, 2, .., l$.

Consider a cycle $\sigma = i_1, i_2, j_2, j_3, i_3, i_4, ..., j_l, j_1$ and the sequence of distances $d_1 = d_L(i_1, i_2)$, $d_2 = d_L(j_2, j_3)$, $d_3 = d_L(i_3, i_4), ....., d_l = d_L(j_l, j_1)$.

The summary distance of the cycle $\sigma$ is defined in [3] as

$$d_{sum,\sigma} = \sum_{i=1}^{l} d_i \quad (2)$$

The summary distance of the permutation can be defined as the minimum of the summary distance of all possible cycles in the permutation, i.e.

$$d_{sum} = \min_{\sigma}(d_{sum,\sigma}) \quad (3)$$

The construction of a graph corresponding to an interleaver is described in [1]. An interleaver graph (IG) for $N = 9$ vertices is shown in Fig. 1 for the permutation $\pi$ given by $\pi = [4\ 7\ 1\ 5\ 8\ 2\ 6\ 0\ 3]$. The edges marked by solid lines in the upper and lower portion of the IG, can be termed as 'upper chain' and 'lower chain' respectively. The edges joining the upper and the lower chains, shown by dotted lines, represent the interleaving function. We call them 'dotted edges'.

In general, we join the vertices marked as 0 and $N-1$ with an edge both in the upper and lower chain of the IG, as $d_L(0, N-1) = 1$, which is a slight modification of the IG of [1]. It can be seen that the IG is a 3-regular graph on $2N$ vertices. The upper and the lower chains are two cycles of length $N$. Other than these two cycles any other cycle in the graph corresponds to a cycle in the permutation $\pi$ as defined earlier and vice versa.

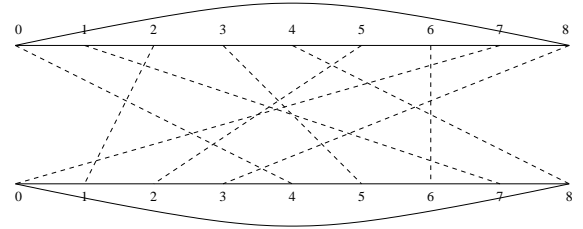

Fig. 1. The modified IG for the permutation example given in Section II

## III. SUMMARY DISTANCE AND GIRTH OF THE IG

Let us call the length of a cycle in the IG, corresponding to a cycle $\sigma$ in the permutation, as $L(\sigma)$ and the number of dotted edges in that cycle as $D(\sigma)$. Clearly the minimum value of $D(\sigma)$ is 2, it is always even, and the maximum value is $\frac{L(\sigma)}{2}$. We can write the summary distance of a cycle as the number of solid edges in a cycle, i.e., $d_{sum,\sigma} = L(\sigma) - D(\sigma)$.

So from (3) the summary distance of an interleaver is

$$d_{sum} = \min_{\sigma}(L(\sigma) - D(\sigma))$$
$$=> \frac{\min_{\sigma}(L(\sigma))}{2} \le d_{sum} \le \min_{\sigma}(L(\sigma)) - 2 \quad (4)$$

where, $\min_{\sigma}(L(\sigma))$ is the girth of the graph.

An upper bound on the girth of a 3-regular graph can be easily obtained and is given in [4]. Using that we can write,

$$\min_{\sigma} (L(\sigma)) \leq 2\log_2(2N+2) - 2 = 2\log_2(N+1) \quad (5)$$

The summary distance of the interleaver, along with the encoders, determines the minimum distance of a turbo code [3]. From (4) and (5) we can conclude that to maximize the summary distance we need to maximize the girth of the IG, which is bounded above by $2\log_2(N+1)$.

## IV. Interleavers from 3-regular Hamiltonian Graphs

In [1] a 4-regular graph is used to construct the IG. For the same number of vertices a 3-regular graph can achieve a higher girth than a 4-regular graph ( [4], p. 180). In the conclusion of [1] it was mentioned that an IG can be constructed from a 3-regular graph if a Hamiltonian cycle can be found in that. But finding a Hamiltonian cycle from a graph is NP-complete. We propose a new construction of interleavers from 3-regular graphs by specifying the Hamiltonian cycle first, then making it 3-regular in a way so that its girth is maximized.

**Construction of IG:** Let the graph $G$ have $N$ vertices which we label 0 to $N-1$ and place them on a cycle, i.e., $i^{th}$ vertex is adjacent to $(i+1) \mod N^{th}$ vertex for $0 \leq i \leq N$. If $G$ is a 3-regular graph then there should be one more edge connected to each vertex. For the moment we assume that the remaining edges are connected arbitrarily. Now the initially fixed cycle becomes the Hamiltonian cycle of the graph. For example in Fig. 2(a) we started with 8 vertices, labeled them 0 to 7, and placed them on a cycle. Then the remaining edges are connected. The resultant graph has an explicit Hamiltonian cycle (the circle in Fig. 2(a)).

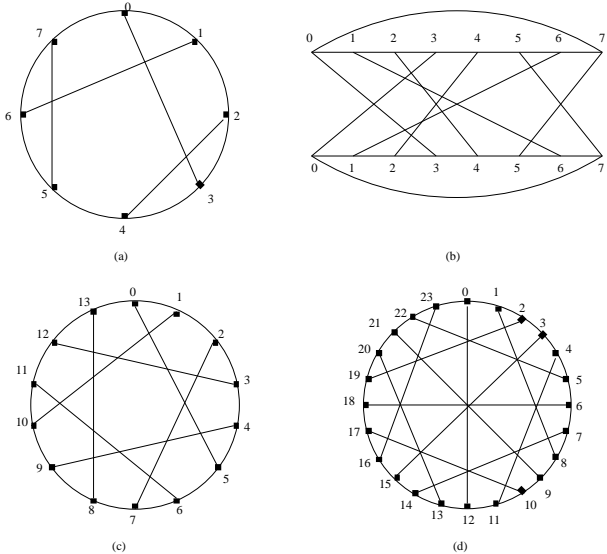

Fig. 2. (a) a 3-regular Hamiltonian graph, (b) the IG constructed from the graph of (a), (c) $14_2(5,7)$ graph, (d) $24_3(12,7,17)$ graph

An IG can be constructed from $G$ in the following way. First draw the upper and the lower chain of the IG with $N$ vertices. Now say $i$ is a vertex from the upper chain and $j$ is a vertex from the lower chain. Connect $i$ and $j$ in the IG, if there is an edge between $i$ and $j$ in $G$ that is not a part of the Hamiltonian cycle of $G$. For example, Fig. 2(b) shows the IG constructed from the graph of Fig. 2(a).

**Proposition 1:** Iff an interleaver $\pi$ is constructed from the 3-regular Hamiltonian graph in the above way then the inverse permutation $\pi^{-1} = \pi$ and $\pi(i) \neq (i \pm 1) \mod N, \forall\ 0 \leq i \leq N-1$.

**Proof:** It can be noticed that, in the interleaver such constructed, if $\pi(i) = j$, then $\pi(j) = i$, otherwise the 3-regularity of the graph is not maintained. This implies $\pi^{-1} = \pi$. As there can be only one edge between two vertices in the graph $\pi(i)$ cannot be $(i \pm 1) \mod N$. Further if any permutation $\pi$ is given where $\pi^{-1} = \pi$ and $\pi(i) \neq (i \pm 1) \mod N$, we can construct a 3-regular Hamiltonian graph from that.

It can be seen that every cycle of the IG constructed from $G$ can be mapped back to a cycle in $G$. So the girth of the IG is lower bounded by the girth of $G$. Thus if $G$ is a 3-regular Hamiltonian graph with large girth, then the corresponding IG will have high girth too.

In the graph $G$, other than the edges which are part of the outer Hamiltonian cycle, the remaining edge connected to a vertex should be selected in a way so that the girth of $G$ becomes high. In [2] a simple construction is proposed which says that the remaining edge incident on $i^{th}$ vertex is determined by a vector $(c_0, ..., c_{s-1})$. The $i^{th}$ vertex is joined to $(i + c_{i \mod s})^{th}$ vertex. The resulting graph is represented using the notation $N_s(c_0, ..., c_{s-1})$.

For example, Fig. 2(c) shows the $14_2(5,7)$ graph and Fig. 2(d) shows the $24_3(12,7,17)$ graph. It can be seen that interleavers can be constructed using only $s$ number of parameters. The interleaver function $\pi$ can be given as

$$\pi(i) = (i + c_{i \mod s}) \mod N \quad \text{for } 0 \leq i \leq N-1 \quad (6)$$

The authors of [5] have employed this technique to construct 'wheel codes'. They have termed the vector $(c_0, ..., c_{s-1})$ associated with a 3-regular Hamiltonian graph as the 'spoke vector'. By proper choice of spoke vectors a graph with high girth can be constructed.

## V. SEARCH FOR SPOKE VECTORS

The number of vertices of a 3-regular graph is always even. So only interleavers of even size can be constructed in this way. For a spoke vector to be valid, the necessary and sufficient conditions are given in [5] as,

$$N \mod s = 0 \quad (7)$$

$$c_i = N - c_{(i+c_i) \mod s} \quad \forall\ 0 \leq i \leq s-1 \quad (8)$$

**Proposition 2:** In a valid spoke vector $c_l \mod s = 0$ if and only if $c_l = \frac{N}{2}$, where $l \in 0, 1, .., s-1$.

**Proof:**

$$c_l = \frac{N}{2}$$
$$<=> \quad c_l = N - c_l$$
$$<=> \quad (l + c_l) \mod s = l \quad \text{from (8)}$$
$$<=> \quad c_l \mod s = 0$$

**Proposition 3:** If $s$ is chosen in a way that (7) is satisfied, the exact number of valid spoke vectors of size $s$ generating a 3-regular Hamiltonian graph of order $N$ is given by

$$\sum_{k=0}^{\frac{s}{2}} \binom{s}{2k} (\frac{N}{s})^{\frac{s}{2}-k}(1.3...(s-2k-1)) \quad s \text{ even}$$

$$\sum_{k=0}^{\frac{s-1}{2}} \binom{s}{2k+1} (\frac{N}{s})^{\frac{s-1}{2}-k}(2.4...(s-2k-1)) \quad s \text{ odd} \quad (9)$$

**Proof:** The proof follows from simple counting arguments. We want to find out the number of possible solutions for the following equations.

$$c_0 = N - c_{c(0) \mod s}$$
$$c_1 = N - c_{(1+c(1)) \mod s}$$

and so on. The last equation is given by,

$$c_{s-1} = N - c_{(s-1+c(s-1)) \mod s} \quad (10)$$

Let us first take the case when $s$ is even. Choosing any $c_i$ will fix $c_{(i+c_i) \mod s}$. So there must be even number of $c_i$ s, which equal $\frac{N}{2}$. Say exact $2k$ elements are equal to $\frac{N}{2}$. All the rest $s-2k$ elements of spoke vectors cannot be a multiple of $s$ (from Proposition 2). They have to be chosen from a set of $N - \frac{N}{s}$ numbers. Choosing any one fixes one more, and the $j^{th}$ choice must be such that $(j + c_j) \mod s \neq (i + c_i) \mod s$, if $c_i$ was chosen before. Suppose this choice can be made in $F_{N,s}(k)$ ways. Then,

$$F_{N,s}(k) = (N - \frac{N}{s})(N - \frac{3N}{s})...(N - \frac{(s-2k-1)N}{s})$$
$$= (\frac{N}{s})^{\frac{s}{2}-k}(1.3.5...(s-2k-1)) \quad (11)$$

So the number of possible solutions is

$$\sum_{k=0}^{\frac{s}{2}} \binom{s}{2k} F_{N,s}(k) \quad (12)$$

Similarly the formula for odd $s$ can also be derived.

We can see from Proposition 3 that the size of the search space for valid spoke vectors is $O(N^{\frac{s}{2}})$. We have to pick the one which is giving the highest girth among them. For a 3-regular Hamiltonian graph with girth $g$, we need to take the size of spoke vector $s \geq \frac{g-2}{2}$. This is due to the fact that there is a cycle of length $2s + 2$ in the graph thus constructed.

We can also generate a spoke vector for a blocklength $N_G$ without searching i.e. using a previously found spoke vector of blocklength $N < N_G$, by slightly generalizing a construction given in [5]. If a description $N_s(c_0, ..., c_{s-1})$ is given then we can construct a graph on $N_G = N + k.s$ vertices if $s$ is even or on $N_G = N + 2k.s$ vertices if $s$ is odd for all $k \geq 1$. If in the initial spoke vector any $c_l + c_m = N$, then in the description $c_l$ is kept unchanged but $c_m$ is written as $-c_l$. In the derived spoke vector make $c_l = \frac{N_G}{2}$ if in the initial spoke vector $c_l \mod s = 0$, else keep $c_l$ unchanged. In the deroved spoke vector $-k$ will represent $N_G - k$.

It is clear that the resultant graph of order $N_G$ with the derived spoke vector will be a 3-regular Hamiltonian graph as $N_G$ is even and (7) and (8) are satisfied.

**Proposition 4:** If the girth of the initial graph on $N$ vertices is $g$ then the girth of the graph on $N_G$ vertices derived in the above way is at least $g$.

**Proof:** We can check for the derived graph $G_s$ for $N + s$ vertices. From the initial to the derived spoke vector the magnitude of the element of spoke vectors and the difference between any two of them is not decreased. If the girth of the derived graph is less than $g$ then there is a cycle in $G_s$ with length less than $g$. It can be shown that a cycle with same or smaller length must be present in the initial graph, leading to a contradiction. Thus the proposition is proved.

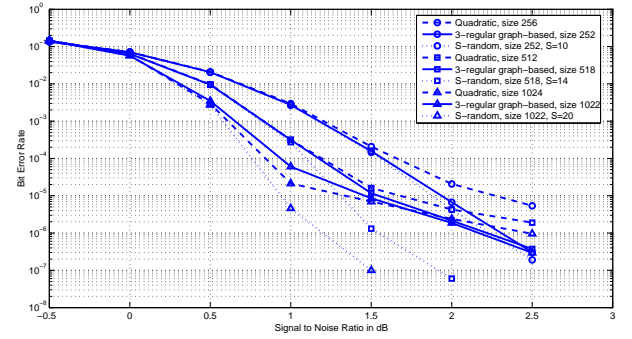

Fig. 3. BER performance of a parallel concatenated convolutional code (PCCC) using MAP decoding (10 iterations).

## VI. SIMULATION RESULTS AND CONCLUSION

We compare the bit error rate (BER) performance of the proposed interleaver with quadratic and S-random interleavers. In [6] it was shown that quadratic interleavers achieve the average performance of random interleavers. Fig. 3 shows the simulated BER performance of a rate $\frac{1}{3}$ PCCC with $(1, \frac{15}{13})$ constituent encoders using BPSK modulation on an additive white Gaussian noise channel. In the high SNR region the graph-based interleaver (solid line) clearly outperforms the quadratic interleaver (dashed line) for all the blocksizes. However, S-random interleavers (dotted line) with high spread outperform both graph-based and quadratic interleavers. But unlike random and S-random interleavers, the proposed graph-based interleavers can be easily generated at both the ends, using a few parameters, and the entire interleaver need not be stored. Further, the interleaver itself works as a de-interleaver.